\documentclass[a4paper]{jpconf}
\usepackage{graphicx}
\begin{document}

\newcommand{\kms}{\mbox{ km s$^{-1}$}}
\newcommand{\be}{\begin{equation}}
\newcommand{\ee}{\end{equation}}
\newcommand{\apj}{ApJ}
\newcommand{\apjs}{ApJS}
\newcommand{\mnras}{MNRAS}
\newcommand{\aap}{A\&A}
\newcommand{\araa}{ARA\&A}
\newcommand{\apjl}{ApJL}
\newcommand{\aj}{AJ}
\newcommand{\nat}{Nature}

\def\ltsima{$\; \buildrel < \over \sim \;$}
\def\simlt{\lower.5ex\hbox{\ltsima}}
\def\gtsima{$\; \buildrel > \over \sim \;$}
\def\simgt{\lower.5ex\hbox{\gtsima}}
\newcommand\sgra{Sgr~A$^*$}
\newcommand\medd{\dot{M}_{\rm Edd}}
\newcommand\mcapt{\dot{M}_{\rm capt}}
\newcommand\ledd{{L}_{\rm Edd}}
\newcommand\mdot{\dot{m}}
\newcommand\Mdot{\dot{M}}
\def\del#1{{}}
\def\degs{$^\circ $}
\def\msun{{\,{\rm M}_\odot}}
\newcommand\mbh{{\,{\rm M}_{\rm bh}}}
\def\rsun{{\,R_\odot}}
\def\lsun{{\,L_\odot}}

\title{Growing supermassive black holes: sub-grid modelling and intermediate-scale processes}

\author{Alexander Hobbs}

\address{Institute for Astronomy, ETH Zurich, Switzerland}

\ead{ahobbs@phys.ethz.ch}

\begin{abstract}
The sheer range of scales in the Universe makes it impossible to model all at once. It is necessary, therefore, when conducting numerical experiments, that we employ sub-resolution prescriptions that can represent the scales we are unable to model directly. In this article we present a prescription for black hole growth that incorporates a different accretion regime from the standard approach used in the literature, and discuss the results of dedicated simulations of intermediate processes between small-scale accretion flows and large-scale cosmological volumes that can strongly enhance the accretion rate onto the black hole at the centre of a galaxy.
\end{abstract}

\section{Introduction}

Modelling galaxy formation requires at least $10$ orders of magnitude in scale, as well as a huge variety of physical processes. These range from the physics of relativistic accretion flows around black holes ($\simlt 10^{-4}$ pc) all the way to gravitational dynamics on the scales of galaxy clusters ($\simgt$ Mpc). In-between there is a host of baryonic physics to include - the hydrodynamics of the gas, radiative cooling, star formation, and magnetic fields, as well as any associated feedback processes. Naturally, all of these couple to each other in complex ways, making the task of simulating galaxy formation extremely difficult.\\

\noindent It is, nonetheless, vital that we do so, for numerical simulations are our only practical means of experimenting with the Universe. The basic procedure for these experiments is relatively simple, consisting of 5 distinct steps:

\begin{enumerate}
\item Build models of what we are interested in
\item Simulate the time evolution of those models subject to physical laws / processes
\item Vary the parameters of the model
\item Analyse the effect on the evolution of the simulation
\item Compare to observations and refine the model
\end{enumerate}

\noindent where this process is repeated to achieve the required level of accuracy and as more detailed observations emerge. The largest stumbling block in this sequence is step (ii), as it is impossible to include all of the required physical processes (even the ones that we know about) due to limitations on computational resources and an inability to resolve all of the necessary scales in a simulation. We must therefore employ prescriptions that encapsulate processes occurring below the resolution limit; so-called ``sub-grid'' models.\\

\noindent One particular process that suffers from these difficulties is the accretion of gas onto a supermassive black hole (SMBH). This is a hugely important aspect of galaxy formation and evolution, as it is now well established that
many galaxies in the local Universe harbour SMBHs
with masses $10^6 \simlt M_{\rm bh}/{\rm M}_{\odot} \simlt 10^9$ at their
centres. Furthermore, SMBHs show a clear relationship with their host galaxy, their masses known to be correlated with observable properties of the galactic bulge. The
best established correlations are between the SMBH mass, $M_{\rm bh}$, and the
velocity dispersion, $\sigma$, i.e., the $M_{\rm bh}-\sigma$ relation
\cite{Gebhardt00, Ferrarese00}, and the $M_{\rm bh}-M_{\rm bulge}$ relation
\cite{Magorrian98,Haering04}. Understanding the formation and growth mechanisms of SMBHs is therefore
crucial in determining the evolution of the larger host systems.

\begin{figure}[h]
\includegraphics[width=25pc]{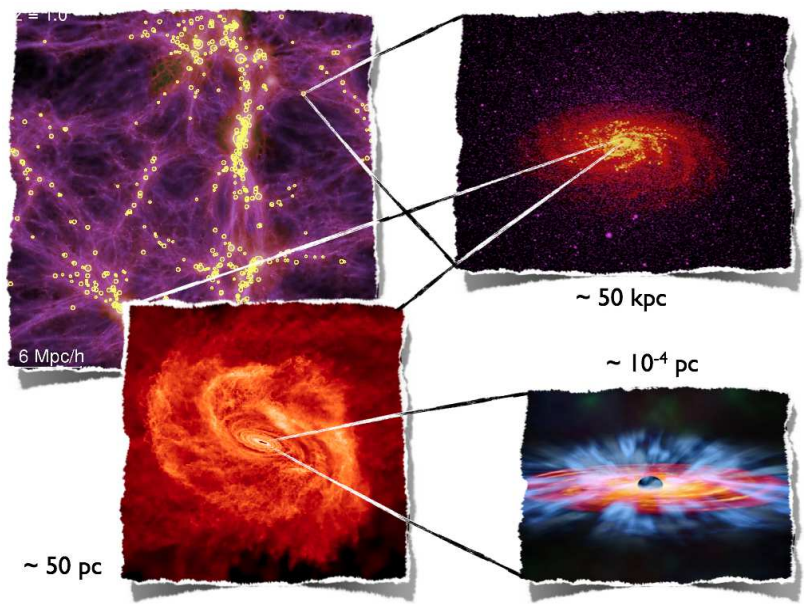}\hspace{2pc}
\begin{minipage}[b]{11pc}\caption{\label{fig:scales} The range of scales required to fully model SMBH growth over cosmic time. Filamentary structures on cosmological scales (top left) funnel gas into forming galaxies (top right), the centres of which harbour SMBHs surrounded by swirling tori of gas (bottom left). As the gas makes its way down to the last stable orbit of the SMBH it forms a dense, hot accretion disc that gives rise to powerful outflows (bottom right). Top left-hand figure from \cite{DiMatteoEtal08}.}
\end{minipage}
\end{figure}

\vspace{0.02in}
\noindent Figure \ref{fig:scales} depicts the range of scales that one must take account of in order to build a complete picture of SMBH growth from the early Universe to the present day. 
It is clear that this growth is a dynamic process, as surveys of active galactic nuclei (AGN) have uncovered the
existence of quasars of mass $M_{\rm bh} \sim 10^9 \msun$ at $z \sim 7$, when the Universe was $\simlt 1/10^{\rm th}$ of its current age;
this implies that many SMBHs had already assembled their mass by this time \cite{MortlockEtal2011}.\\

\noindent Of course, our understanding of how SMBHs grow to these masses is
incomplete. We know that black holes grow by accreting
low angular momentum material from their surroundings, yet the character
of the accretion flow onto an SMBH is governed by physical
processes as diverse as galaxy mergers \cite{HopkinsQuataert2010},
turbulence induced by stellar feedback \cite{HobbsEtal2011} and
black hole accretion-driven outflows \cite{nayakshin.power.2010}.\\

\noindent The typical masses of ``seed'' BHs in the early Universe are under debate, but it is still clear that to reach the required $\sim 10^9 \msun$ the SMBH must have experienced a high and sustained accretion rate. Even at the higher end of possible seed masses, i.e., $\sim 100 \msun$ \cite{MadauRees01}, we would still require Eddington-limited growth in order to reach the observed masses of early Universe quasars \cite{KingPringle07}. One must therefore conclude that the accretion flow, as well as the evolution of the galaxy as a whole, is likely strongly influenced by AGN feedback that is coupled to the SMBH growth rate \cite{PowerEtal2011}.


\section{Sub-grid modelling of SMBH growth}

Black hole growth is now routinely modelled in galaxy formation simulations
(for a fiducial work see \cite{SpringelEtal05}) and the importance of SMBHs in shaping the
properties of galaxies has been demonstrated \cite{croton.etal.2006,bower.etal.2006,DiMatteoEtal08}. Given the difficulty in accounting for the sheer range of scales involved, we are compelled to make use of sub-grid modelling. The majority of galaxy
formation simulations published in the literature incorporate what we shall
term the ``Bondi-Hoyle model'' for black hole growth
\cite{SpringelEtal05, SijackiEtal2007, PelupessyEtal2007, DiMatteoEtal2008, JohanssonEtal2009, KimEtal2011}, which derives from the work of \cite{Bondi44}
and \cite{Bondi52} - hereafter B\&H. This model assumes the simplest possible accretion flow,
where the gas is at rest at infinity and accretes steadily onto a black hole,
subject only to the (Newtonian) gravity of the latter, which is modelled as a
point mass.\\

\noindent In galaxy formation
simulations, unfortunately, this idealised picture is far from satisfied, as the gas inflow is complicated considerably by the properties of the flow at larger scales. Owing to the requirement of modelling the inflow and eventual accretion in a sub-grid fashion, we find it is useful to cast the infalling gas into several accretion ``regimes'', with the relative contribution each determined from the character of the flow at the resolution limit of the simulation. To start with, we have the two extremes:

\begin{enumerate}
\item stationary, where the gas is completely supported and at rest at the resolution limit
\item in free-fall, where the gas flow is entirely radial and influenced only by gravity
\end{enumerate}

\noindent It is important to recognise that the reality will lie somewhere in between. Specific cases that lie between these extremes are:

\begin{itemize}
\item the gas is unsupported at the resolution limit but shocked at smaller radii, after which it may begin to tend to free-fall once again if the cooling time is short
\item the gas possesses angular momentum, and is thus supported by rotation but inflows through viscous processes
\end{itemize}

\noindent The mode of Bondi-Hoyle accretion corresponds to option (i), and so employing the Bondi-Hoyle sub-grid prescription in simulations implicity assumes such a regime. The opposite case, that of the gas being in free-fall, is therefore not catered for in this standard model.

\subsection{The Bondi-Hoyle accretion regime}

The nature of the Bondi-Hoyle regime for accretion is summarised in the first sentence of the
\cite{Bondi52}'s abstract: ``The special accretion problem is investigated
in which the motion is steady and spherically symmetrical, the gas being at rest at infinity''. Physically, gas can only be at rest if it is not subject to any forces. The only external force acting on the gas in the B\&H problem, i.e., the gravitational force, is due to
the black hole. Self-gravity of the gas is neglected. The ``infinity'' in question is a
region at a distance large enough from the SMBH that the gravitational force exerted by the latter
is negligible when compared to the pressure forces within the gas. This is quantified by
defining the Bondi radius,
\begin{equation}
r_B = 2GM_{\rm bh}/c_\infty^2
\end{equation}
where $M_{\rm bh}$ is the mass of the central object and $c_\infty$ is the
sound speed of the gas far from the hole.\\

\noindent Applying the Bondi-Hoyle formalism to black hole growth assumes that the
accretion rate onto the SMBH is commensurate with the accretion rate through
the Bondi radius (i.e., the flow is steady-state) and therefore given by
\begin{equation}
\dot M_{\rm bh} = \pi \lambda(\Gamma) r_B^2 \rho_\infty c_\infty = \frac{4 \pi
  \lambda(\Gamma) G^2 M_{\rm bh}^2 \rho_\infty}{c_\infty^3}
\label{eq:mdt}
\end{equation}
where $\rho_\infty$, $c_\infty$ are the density and sound speed at infinity respectively, and $\lambda(\Gamma)$ contains all the corrections arising due to the finite
pressure gradient force in the problem. In sub-grid modelling, however, the relevant gas properties ``at infinity'' are evaluated at a radius around the SMBH that corresponds to the smallest resolvable scale. Often this is orders of magnitude larger than the Bondi radius. We clearly require the gas to be at rest at this evaluation radius, and for the SMBH to dominate the gravitational potential in order for the Bondi-Hoyle formalism to be applicable, even to spherical flow. Such a situation could be achieved if the gas within the SMBH sphere of influence is at the virial temperature and therefore in hydrostatic balance \cite{KomatsuSeljak2001, SutoEtal98}. This may be the case for low-luminosity SMBHs in giant elliptical
galaxies where the gas is rather tenuous and hot due to a long cooling time \cite{ChurazovEtal2005}.

\subsection{The free-fall accretion regime}

However, in cosmological simulations (as in the real Universe) we expect SMBHs to be immersed in stellar bulges and dark matter haloes that are typically $\sim 10^3$ to $10^4$ times
more massive than the SMBH \cite{Haering04, GuoEtal2010}. Furthermore, in the epoch when SMBHs grow rapidly, their hosts are very gas
rich, and the inflow of gas from large scales cannot be easily captured by gradual cooling from a tenuous hot halo \cite{BirnboimDekel2003, KeresEtal2005,
KeresEtal2009, DekelEtal2009, KimmEtal2011}. Higher density gas is likely to cool much faster and hence is
likely to be much cooler than the virial temperature. We therefore expect an inflow of gas to
the centre rather than hydrostatic balance.\\

\noindent Due to the dominant potential energy of the halo, and the fact that the temperature of the gas infalling from large scales cannot be greater than the virial temperature at the halo virial radius, we can see that for spherical flow with a relatively soft equation of state the character of the inflow may reach that of free-fall at the evaluation radius instead of being stationary.  

\subsection{An interpolation approach}

Of course, the two regimes discussed above constitute a simplified picture. In reality gas that has accreted onto a halo from outside the virial radius may shock at smaller radii, heating up to the local $T_{\rm vir}$ at that radius. However, due to the high densities reached for the shocked gas the cooling time is likely to be short \cite{BirnboimDekel2003, KeresEtal2005}, and in particular significantly less than the free-fall time \cite{NulsenFabian2000}. Gas may therefore be stationary temporarily at the shock radius but as it cools and begins to infall it will quickly tend to the free-fall velocity, and certainly by the time a radial inflow reaches the Bondi radius the assumption of being at rest will no longer be satisfied.\\

\noindent The list of regimes above is essentially a range of ``supporting mechanisms'' for the gas at the evaluation radius, the relative importance of which must be taken account of to infer the appropriate accretion rate. This is very difficult to do in a simulation, and so as a starting point we deal solely with the two extremes - (i) and (ii) as referred to above. We require a formula that interpolates between these two cases, and so we propose, as per \cite{HobbsEtal2012}, the expression
\begin{equation}
\dot M_{\rm interp} = \frac{4 \pi \lambda(\Gamma) G^2 M_{\rm enc}^2 \rho_\infty}{(c_\infty^2 + \sigma^2)^{3/2}}
\label{eq:minterp}
\end{equation}
where $\sigma$ is the velocity dispersion for the external potential, $\sigma \sim (GM_{\rm enc}(r)/r)^{1/2}$ and $M_{\rm enc}$ is the enclosed mass. This expression tends to the standard Bondi-Hoyle formula (equation \ref{eq:mdt}) in the limit that $c_\infty \gg \sigma$, i.e., the gas internal energy dominates over the potential energy of the halo, and $M_{\rm enc} \rightarrow M_{\rm bh}$, i.e., as we approach the SMBH radius of influence. In the opposite limit, where the halo potential energy dominates, $\sigma \gg c_\infty$, we recover the free-fall estimate, $M_{\rm gas}/t_{\rm ff}$, where $M_{\rm gas}$ is the enclosed gas mass and $t_{\rm ff} = (r^3/2GM(r))^{1/2}$ is the free-fall timescale. This interpolation approach is a minor modification of the full Bondi-Hoyle-Lyttleton \cite{HoyleLyttleton1939, Bondi52} expression for an accretor that is moving relative to the gas, with two changes: (1) the velocity dispersion replaces the relative velocity between gas and accretor, and (2) the total enclosed mass (external potential + gas + SMBH) replaces the black hole mass. Finally, in order to incorporate (in a crude fashion) the effect of AGN feedback on the accretion the rate should be capped at Eddington (as is commonly employed in simulations), namely
\begin{equation}
\dot M_{\rm acc} = \texttt{max}(\dot M_{\rm interp}, \dot M_{\rm Edd})
\end{equation}


\section{Turbulent feeding on intermediate scales}

A major caveat which we must point out with the discussion so far is that the presence of angular momentum in the gas alters things significantly. Indeed, angular momentum can be viewed as a physical ``barrier'' to accretion, with only the lowest angular momentum material able to flow to smaller scales and accrete onto the SMBH. One should therefore endeavour to take account of angular momentum in a sub-grid model of SMBH growth if at all possible. For this we recommend the ``accretion disc particle'' method of \cite{PowerEtal2010}, which tracks and utilises the angular momentum in the gas to infer the rate of accretion and feedback.

\begin{figure}[h]
\includegraphics[width=24pc]{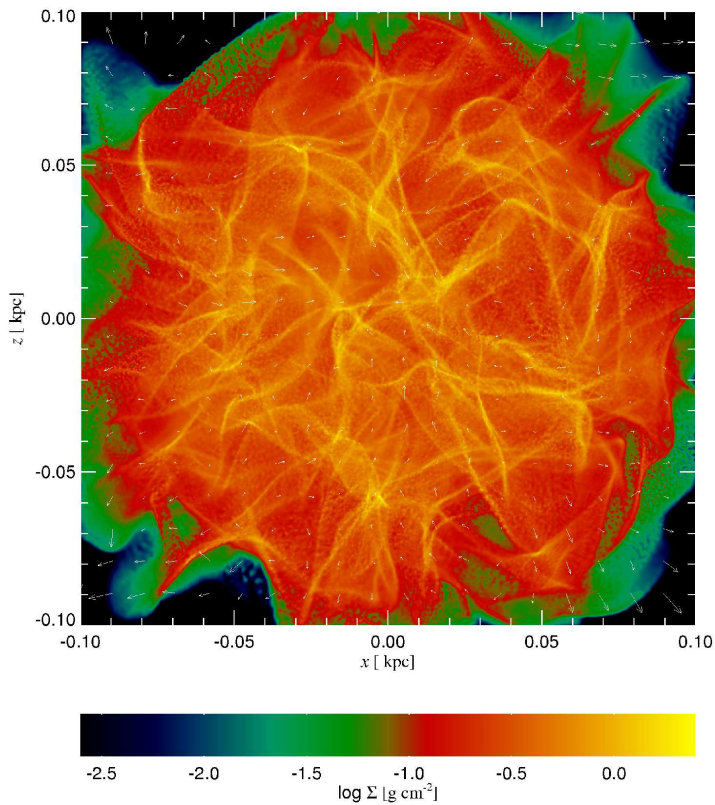}\hspace{2pc}%
\begin{minipage}[b]{12pc}\caption{\label{fig:projection}Projected gas column density in the fiducial turbulent feeding simulation, at an early time ($10^5$ yrs),
  before gas has had a chance to accrete on the SMBH. Note the formation of multiple
  thin and dense filaments due to convergent turbulent velocity flows.}
\end{minipage}
\end{figure}

\noindent Of course, taking account of angular momentum in the sub-grid model will only make it harder for an SMBH to accrete, as viscous timescales, even at parsec scales, are often prohibitively long (with standard assumptions \cite{Shakura73}, $t_{\rm visc}$ through a $\sim 10$ pc scale disc is $\sim t_{\rm Hubble}$). We therefore need a process that can allow gas to acquire low angular momentum. A promising candidate is the supernovae-driven turbulent feeding model of \cite{HobbsEtal2011}, which would typically occur at scales intermediate between detailed simulations of black hole accretion discs and large-scale cosmological simulations, and is thus not captured by the majority of galaxy formation models. Indeed, it is important to take account of processes occurring on intermediate scales which may often get missed by this dichotomy of approaches to SMBH growth.\\

\noindent Figure \ref{fig:projection}, taken from \cite{HobbsEtal2011}, shows a visualisation from a numerical simulation of a mode of turbulent feeding at scales of $\sim 100$ pc. The turbulent velocity field imposed on the gas (in addition to a net rotation) is tailored to the typical energy and distribution of supernovae explosions in a spherical shell of gas at these scales. The formation of overdense filaments can clearly be seen, due to the action of the supernovae feedback in sweeping up and shocking the gas.\\

\begin{figure}[h]
\includegraphics[width=22pc]{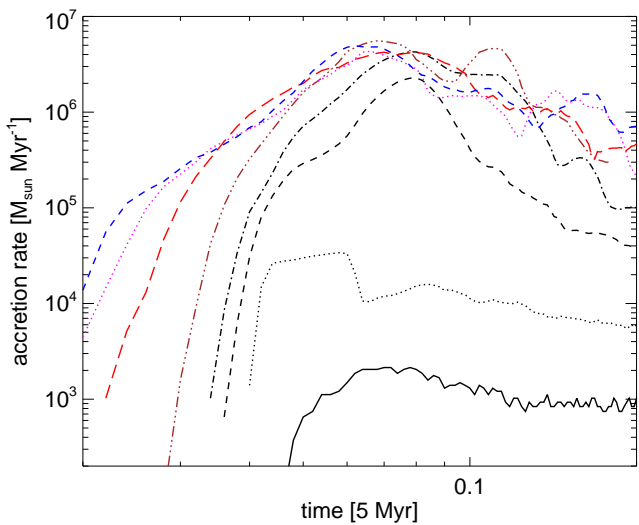}\hspace{2pc}%
\begin{minipage}[b]{14pc}\caption{\label{fig:accretionrate}Accretion rate versus time for a fiducial set of simulations, with $v_{\rm rot} = 60$ km s$^{-1}$. Linestyles correspond to $v_{\rm turb} = 0$ (black solid), $20$ km s$^{-1}$ (black dotted), $40$ km s$^{-1}$ (black dashed), $60$ km s$^{-1}$ (\
black dot-dash), $100$ km s$^{-1}$ (brown dot-dot-dash), $200$ km s$^{-1}$ (red long dashed), $300$ km s$^{-1}$ (pink dotted), and $400$ km s$^{-1}$ (blue dashed) in code units. We can clearly see that the accretion rate on the black hole strongly increases with increasing levels of turbulence when rotation is present.}
\end{minipage}
\end{figure}

\noindent This turbulent feeding model yields some very interesting and potentially useful results in terms of the accretion rate through the central $1$ pc (the accretion radius, $r_{\rm acc}$, of the simulations). There is a strong trend of enhanced accretion with levels of increasing turbulence - indeed, with a mean turbulent velocity of $\sim 200$ km s$^{-1}$, which is the approximate value of the velocity dispersion at these scales, the accretion rate is $2-3$ orders of magnitude higher than in the same simulation with no seeded turbulence. These results can be seen in Figure \ref{fig:accretionrate}. The explanation for this lies in the angular momentum distribution with and without turbulence. When the velocity field was purely rotational (about a single axis), the angular momentum distribution was highly peaked, corresponding to a thin disc or a ring at a particular radius outside $r_{\rm acc}$. This ring can only accrete into the central parsec by viscous processes, thereby yielding a very low accretion rate. With the introduction of a turbulent velocity field, however, the angular momentum distribution was significantly broadened, putting some of the gas on low enough angular momentum orbits to travel inside the accretion radius \emph{ballistically}, unaffected by hydrodynamical drag due to the high overdensity of the infalling gas clump or filament.\\

\noindent The dominant mode of feeding is therefore a ballistic one, but to later times an extended (and still quite turbulent) disc forms which must then accrete viscously. Such a structure is however no longer a ring but instead extends over a wider range of radii and accretes more rapidly than the non-turbulent case - this is due to the action of a supersonic turbulent viscosity that is maintained over the lifetime of the infall, as gas from the outer parts of the shell continues to impact upon the disc. Turbulence thus enhances the accretion rate in both modes of feeding. Of course, if the velocity field had no rotational component the opposite effect would be seen, since without turbulent broadening of the distribution all of the gas would simply end up inside the accretion radius. As we have mentioned, however, the gas typically inflows from considerably larger scales and so it is likely that angular momentum will, by the time the gas has reached these intermediate scales, translate into significant rotation.\\

\noindent The supernovae-driven turbulent feeding model can be described analytically by convolving a distribution of turbulent velocities (typically Maxwellian) with the mass of gas in the loss-cone of an isotropically-emitted wind in rotation around the SMBH - see \cite{HobbsEtal2011} for more details. For a shell of initially constant density, with the outer radius $\approx 3$ $\times$ the inner radius, this yields an accreted mass fraction of
\begin{equation}
\frac{\Delta M}{M_{\rm shell}} = \frac{3 (GM_{\rm bh} r_{\rm acc})^{1/2}}{2 \sqrt{2} r_{\rm out} v_{\rm rot}} \cdot
\psi + \frac{9 (GM_{\rm bh} r_{\rm acc})^{1/2}}{4 \sqrt{2} r v_{\rm turb} \pi^{1/2}} \cdot \varphi
\label{eq:mlc_shell}
\end{equation}
where $v_{\rm turb}$ is the mean turbulent velocity, and $v_{\rm rot}$ is the rotation velocity. The functions $\psi$ and $\varphi$ are given by
\begin{displaymath}
\psi \equiv \texttt{erf}\left(\frac{5 v_{\rm rot}}{4 v_{\rm turb}}\right) -
  \texttt{erf}\left(\frac{3 v_{\rm rot}}{4 v_{\rm turb}}\right)
\end{displaymath}
\begin{displaymath}
\varphi \equiv e^{-9 v_{\rm rot}^2 /16                                                                
    v_{\rm turb}^2} - e^{-25 v_{\rm rot}^2 /16 v_{\rm turb}^2}
\end{displaymath}
which, in the limit that $v_{\rm turb} \gg v_{\rm rot}$, simplifies to
\begin{equation}
\frac{\Delta M}{M_{\rm shell}} \approx \frac{(2 GM_{\rm bh} r_{\rm acc})^{1/2}}{r_{\rm out}
  v_{\rm turb}} \left(1 + \frac{v_{\rm rot}^2}{v_{\rm turb}^2}\right)
\label{eq:mlc_gg}
\end{equation}
where we have dropped numerical factors of approximately unity.\\

\noindent This analytical theory reproduces the numerical results from the simulations well \cite{HobbsEtal2011}, and can be used to make a prediction for the surface density of the disc that forms once the main ballistic mode of feeding ends. This prediction is the surface density is quite steep, $\Sigma(R) \propto R^{-3/2}$. We note that \cite{KishimotoEtal2011} have recently found evidence that for high-luminosity AGN, the surface density distribution of heated dust (a tracer for the AGN disc) inferred from mid-IR interferometry can indeed be quite steep, going as $R^{-1} \simlt \Sigma(R) \simlt R^{-3/2}$ (refer to Figure 14 in \cite{KishimotoEtal2011}). The supernovae-driven turbulent feeding model is therefore a promising one both in terms of inferred SMBH accretion rates and observations of AGN tori. Moreover, the implication that the progenitor for the turbulence is feedback from star formation speaks to a connection between the stellar properties of the bulge and the SMBH mass, such as is observed.

\section{Conclusions}

\noindent Sub-grid modelling is a vital aspect of galaxy formation simulations. Processes occurring below the limit of resolution can be encapsulated into sub-grid prescriptions to allow us to explore the formation and evolution of galaxies over cosmic time. It is, however, extremely important that sub-grid models continue to be refined and improved using dedicated simulations as our understanding of the relevant processes improves. In particular, the extension of existing SMBH growth prescriptions to include different accretion regimes will help to shed light on the existence of quasars at high redshift and the observed SMBH-host correlations through the action of feedback. The Bondi-Hoyle/free-fall interpolation model of \cite{HobbsEtal2012} attempts to account for the opposite extreme from the standard Bondi-Hoyle formalism in models of SMBH growth. Taking an approach to sub-grid modelling that is physically motivated, rather than empirically-motivated, will allow numerical simulators to construct models from first principles.\\

\noindent An important area of sub-grid modelling that is often overlooked is that of ``intermediate'' scales, which may be too small to be captured at high resolution in simulations of, say, galaxy mergers, but too large to be of interest to simulators studying detailed accretion flows at the last stable orbit around an SMBH. However, processes occurring on these intermediate scales can strongly influence the flow onto the SMBH, and indeed provide a potential solution to the AGN feeding problem. The supernovae-driven turbulent feeding model of \cite{HobbsEtal2011} may be such a process.

\section{References}

\bibliographystyle{iopart-num}

\bibliography{refs}

\end{document}